\documentstyle[11pt]{article}
\newcommand{\be}{\begin{equation}}
\newcommand{\ee}{\end{equation}}
\newcommand{\bea}{\begin{eqnarray}}
\newcommand{\eea}{\end{eqnarray}}
\newtheorem{definition}{Definition}[section]
\newtheorem{theorem}{Theorem}[section]
\newcommand{\al}{\alpha}
\newcommand{\Ho}{{\cal HG}_{x_0}}
\newcommand{\sub}{\subseteq}
\newcommand{\supsetneq}{{\scriptsize\begin{array}{c}\supset\\ \neq\end{array}}}
\newcommand{\Jo}{J_{x_0}}
\newcommand{\ao}{\al_{x_0}}
\begin{document}
\title{\hfill {\rm\normalsize NBI-HE-94-46} \\ \ \\
Induced Topology on the Hoop Group}
\author{J\o rgen Rasmussen\thanks{jrasmussen@nbivax.nbi.dk} \ and Morten
Weis\thanks{weis@nbivax.nbi.dk} \\ Niels Bohr Institute \\
University of Copenhagen \\ Blegdamsvej 17, DK-2100 Copenhagen \\
Denmark}
\date{October 24, 1994}
\maketitle
\begin{abstract}
A new topology is proposed on the space of holonomy equivalence classes of
loops, induced by the topology of the space $\Sigma$ in which the
loops are embedded. The possible role for the new topology in the context
of the work by Ashtekar et al. is discussed.
\end{abstract}

In this short paper we introduce a topology $\tau$ on the space of holonomy
equivalence classes of loops, known in the context of quantum gravity in
the Ashtekar variables as the hoop group ${\cal HG}$.
The first section is devoted to notation by introducing hoops etc.
The main part is the second section in which the topology is defined.
In the last section we briefly discuss how this topology might be
of interest for the work by
Ashtekar et.al.~\cite{Ashtekar1}~\cite{Almmt}
regarding the non-linear functional analysis of the quotient space of
connections modulo gauge transformations ${\cal A/G}$.

\section{Introduction and Notation}
Consider a $d$-dimensional real analytical oriented paracompact
manifold $\Sigma$.
Let us introduce the notation of continuous,
piecewise analytic parametrized loops in $\Sigma$
\be
   \al:[0,\lambda_{1}]\cup[\lambda_1,\lambda_2]\cup\dots\cup[\lambda_{n-1},1]
       \mapsto \Sigma\ \ \ \,
     \ \ \  \al(0)=\al(1)
\ee
The mappings are continuous on the whole domain and analytical on the
single closed intervals $[\lambda_{i},\lambda_{i+1}]$.

We consider only loops which all share the
arbitrary but fixed point $x_{0}\in \Sigma$. Denote
this set of based loops in $\Sigma$ by ${\cal L}_{x_{0}}$. The standard
composition "$\circ$" of loops makes ${\cal L}_{x_{0}}$ a semigroup.
The principal fibre bundle $P(\Sigma,G)$, where $G$ is a Lie group,
connections and holonomies are defined as usual~\cite{Kob}.
The topology on $\Sigma$ is unspecified
so we cannot say whether the bundle will be trivial or not.
Given a connection $A$ and a based loop $\alpha\in
{\cal L}_{x_{0}}$ we can compute the holonomy element $H(\alpha,A)$,
which takes values in the group $G$. We follow
Ashtekar and Lewandowski~\cite{Ashtekar1} and introduce holonomy
equivalence classes.

\begin{definition}
A pair of loops $\alpha,\beta \in {\cal L}_{x_{0}}$ are holonomy equivalent
$\alpha \sim \beta$ if $H(\alpha,A)=H(\beta,A)$ for all $G$-connections $A$.
The set of equivalence classes $[\al]$ is denoted
${\cal HG}\equiv\Ho = {\cal L}_{x_{0}}/\sim$, the space of hoops.
\end{definition}
{\bf Remark:}
The hoop space ${\cal HG}$ is a group under
\be
  [\al]\cdot[\beta]=[\al\circ\beta]\,\,\,\,\, ,\,\,
       \,\,\,\,\,[\ao]={\bf 1}_{\Ho}
\ee
where $\ao :[0,1]\mapsto x_0$ is the constant loop.

In the next section we propose a topology on the hoop group, which
to our knowledge has not yet been discussed.

\section{The Topology $\tau$}
Let $\tau_{\Sigma}$ denote a fixed topology on $\Sigma$. The open
neighbourhood filter of $x_0$ is denoted by ${\cal O}(x_0)$ and defined to be
the family of open neighbourhoods of $x_0$
\be
  {\cal O}(x_0)=\{ U_j\}_{j\in J_{x_0}}
\ee
where $J_{x_0}$ is an index set depending on $\tau_{\Sigma}$ while
$U_j\in\tau_{\Sigma}$. \\
Now we define the set of "topology generators" {$V_j$} in $\Ho$.
They are induced
by {$U_j$} in the following way
\be
  V_j=\{ [\al ]\in \Ho\ |\ \exists \al'\in [\al ] \ ,\ \al'\sub U_j\}
\label{Vj}
\ee
Some additional notational definitions are introduced
\be
  V_{j_1\cdots j_n}=\{[\al ]\in\Ho\ |\ \exists\al'\in [\al ]\ ,\ \al'\sub
                   \bigcap_{i=1,\cdots,n} U_{j_i}\}
\ee
\be
  V_{(j_i)_{I_j}}=\{[\al ]\in\Ho\ |\ \exists\al'\in [\al ]\ ,\ \al'\sub
                   \bigcup_{i\in I_j} U_{j_i}\}
\label{Vji}
\ee
where $I_j$ is chosen such that for any $i\in I_j$ , $j_i\in J_{x_0}$.
Such an introduction is convenient because generically we have
\be
  V_{(j_1,j_2)}\supsetneq V_{\{j_1,j_2\}}\equiv V_{j_1}\cup V_{j_2}
\ee
and more generally
\be
  V_{(j_i)_{I_j}} \supsetneq V_{\{j_i\}_{I_j}}
    =\{[\al]\in\Ho\ |\ \exists\al',\exists U_{j_i},j_i\in I_j\ ,\ \al'\subseteq
             U_{j_i}\}
\ee
On the other hand the following equation is valid
\be
  V_{j_1\cdots j_n}=\bigcap_{i=1,\cdots,n}V_{j_i}
\label{snit}
\ee
The concept "topology generator" introduced above is accounted
for by our main claim
\begin{theorem}
  $\tau=\tau(\{V_j\}_{j\in J_{x_0}})$  is a topology on $\Ho$ where
   $\{V_j\}_{j\in J_{x_0}}$ is a base of $\tau$ in the usual (topological)
sense.
\end{theorem}
{\bf Proof:}
  We only have to check that $\{V_j\}_{j\in J_{x_0}}$ is stable under finite
intersections. By construction (\ref{Vj}) and (\ref{snit}) this is obvious
\bea
  \bigcap_{i=1,\cdots,n}V_{j_i}&=&\{[\al ]\in\Ho\ |\ \exists\al'\in[\al ]\ ,
   \ \al'\subseteq\bigcap_{i=1,\cdots,n}U_{j_i}\} \nonumber \\
  &\equiv&\{[\al ]\in\Ho\ |\ \exists\al'\in[\al ]\ ,\ \al'\subseteq U_k\}
     \nonumber\\
  &=&V_k \ \ \ \Box
\eea
{\bf Remark:} By this definition the empty set $\emptyset$ is not an open
set, but by the enlargement $\tau'=\tau\cup\emptyset$ we obtain a perhaps
more conventional type of topology\footnote{According to \cite{Kelley} and
others it is strictly speaking not necessary to include $\emptyset$ to
have a topology.}. \\
\noindent
Because $[\ao]\in V_j$ for any $j\in\Jo$, ($\Ho,\tau$) is not a
Hausdorff topological space.

\section{Outlook}

The question of topology on the hoop group is interesting in its own
right, but our motivation has come from the use of the hoop group in the
context of quantum gravity in the Ashtekar formalism. In this
context $\Sigma$ is chosen as a 3-manifold and usually one chooses
$G=SU(2)$ whereby $P(\Sigma,G)$ is trivial. $\Sigma$ plays the role as a
space-slice in the canonical 3+1 dimensional formulation of quantum gravity.
The topology of classical
space-time is chosen to be $\Sigma \times {\bf R}$. The role of the
hoop group is related to the problem of constructing measures on
the configuration space, which is the essence of a quantum field theory.

In the recent work by Ashtekar et.al.~\cite{Ashtekar1}~\cite{Almmt} a
non-linear duality between loops and connections has been explored, in
order to generalize the notation from ordinary QFT~\cite{Jaffe}
to the non-linear
configuration space of Ashtekar gravity and full Yang-Mills theory in
4 dimensions. The holonomy defines this duality and by using ${\cal HG}$ as
the dual space to ${\cal A/G}$, Ashtekar and Lewandowski have introduced
cylindrical measures on ${\cal A/G}$~\cite{Ashtekar1}. The holonomy
can be viewed as a functional $T_{\al}(A)=
tr H([\al],[A]_{{\cal G}})$ on ${\cal A/G}$.

The characteristic function is a functional on ${\cal HG}$~\cite{Almmt}
\be
\chi([ \al]) = \int_{\overline{{\cal A/G}}} \overline{T}_{[ \al]}(
\overline{[A]_{{\cal G}}}) d\mu_{AL}(\overline{[A]_{{\cal G}}})
\ee
where the bars denote that the integral and the $T_{[\al]}$
functional are extended to a larger space,
here the algebraic dual of ${\cal A/G}$.
$d\mu_{AL}$ is the Ashtekar-Lewandowski measure, induced by the
Haar measure on $SU(2)^{n}/Ad$ in the projective limit.
As stated in~\cite{Almmt} the function is only known to
be continuous in the discrete topology on ${\cal HG}$. We see that hoops play
the role as test-functions in this scheme and the topology on the hoop group
plays a role similar to the topology on the Schwarz space in ordinary QFT.
The discrete topology is known to be too simple for the formulation of an
interesting quantum theory. A more promising topology has been introduced by
Barret~\cite{Barret} and discussed by Lewandowski~\cite{Lewandowski},
though one has not yet been able to
use it in the study of continuity of $\chi([\al])$.

It will be interesting to know whether $\chi([\al])$ is continuous or
not in the topology $\tau$ (depending on $\tau_{\Sigma}$)
proposed in this paper.
This and other questions are under current investigation~\cite{jm}.

\section*{Acknowledgements}
One of us (MW) wishes to thank A. Ashtekar, E. Goldblatt and T. Thiemann
for interesting discussions
at the gravity center at Penn State Univ.
MW is grateful for financial support by the
Danish Natural Science Research Council, grant 11-0801-1.


\begin{thebibliography}{99}
\bibitem{Ashtekar1}A. Ashtekar and J. Lewandowski.
{\it Representation Theory of
Analytic Holonomy $C^{*}$-Algebras.} In: Knots and Quantum Gravity, ed. J.
Baez, (1994) Oxford University Press (Oxford).
\bibitem{Almmt} A. Ashtekar, J. Lewandowski, D. Marolf, J. Mour\~{a}o and
T. Thiemann. Preprint CGPG-94/8-2, hepth:9408108.
\bibitem{Kob} S. Kobayashi and K. Nomizu. {\it Foundations of Differential
Geometry}. Vol. 1, (1963) Interscience Publishers (New York).
\bibitem{Kelley}J.L. Kelley. {\it General Topology}. Graduate Texts in
Mathematics vol. 27, (1955) Springer-Verlag (New York).
\bibitem{Jaffe} J. Glimm and A. Jaffe. {\it Quantum Physics - A Functional
Integral Point of View.} 2.ed. edition. (1987) Springer-Verlag (New York).
\bibitem{Barret} J.W. Barret. Int. J. Theo. Phys. {\bf 30} (1991) 1171;
\bibitem{Lewandowski} J. Lewandowski. Class. Quant. Grav. {\bf 10} (1993) 879;
\bibitem{jm} J. Rasmussen and M. Weis. Work in progress.


\end{thebibliography}
\end{document}